\documentclass{mem}
\usepackage{natbib}\usepackage{txfonts}\usepackage{balance}
\usepackage{graphicx}
\usepackage[a4paper,breaklinks,dvipdfm]{hyperref}
\idline{75}{282}
\begin{document}
\def\teff{$T\rm_{eff }$}
\def\kms{$\mathrm {km s}^{-1}$}

\title{
Disk-Jet connection in outbursting Black Hole sources
}

   \subtitle{}

\author{
D.\, Radhika\inst{1,2} 
A.\, Nandi\inst{1}
\and S.\, Seetha\inst{1}
          }

  \offprints{D. Radhika}

\institute{
Space Astronomy Group, ISRO Satellite Centre, Bangalore, INDIA
\and
Department of Physics, University of Calicut, INDIA \\\\
\email{radhikad.isac@gmail.com}
}

\authorrunning{Radhika }

\titlerunning{Disk-Jet connection in outbursting Black Hole sources}

\abstract{
We explored the `spectro - temporal' behaviour of outbursting Black Hole sources in X-rays, at 
the time of jet ejections which are observed as radio flares. The energy dependent evolution of 
the properties of all the sources studied, shows that during the ejections the QPO frequencies 
`disappear' as well as the disk (thermal) emission increases, implying the soft nature of the 
spectrum. These results can be understood based on the TCAF model \citep{CT95}, in the presence 
of magnetic field. 
\keywords{Black holes, Accretion physics, X-ray sources, Radiation hydrodynamics }
}
\maketitle{}

\section{Introduction}

Galactic Black hole (BH) sources are interesting objects to study as the process of accretion 
gets very complex as the disk evolves with time, especially when the sources undergo an 
outburst and Jet ejections take place. It is observed that the peak of the radio emission is 
associated with the hard to soft state transition \citep{FHB09}. 
We study the disk-jet connection of the outbursting BH sources (e.g. XTE J1859$+$226, 
XTE J1752$-$223, XTE J1748$-$288 and H1743$-$322) based on their `spectro-temporal' properties.

\section{Observations and Results}

We analyzed the archival data of RXTE PCA/HEXTE in the energy band of 2 - 150 keV during the 
radio flares (i.e., Jets) of the outbursting BH sources.
We observe an increase in thermal flux and the ratio of soft (3 - 20 keV) to hard (20 - 50 keV) 
flux in all the sources during the ejections, which implies the softening of the spectra. 
Energy dependent study was done to understand the evolution of QPOs, nature of PDS, rms 
etc. during the radio flares, and the interesting features that we observed are summarized 
below. 

\subsection{XTE J1859$+$226: 1999 outburst} 
\begin{figure*}[t!]
\begin{minipage}[t]{6cm}
\includegraphics[height=3cm,width=6cm]{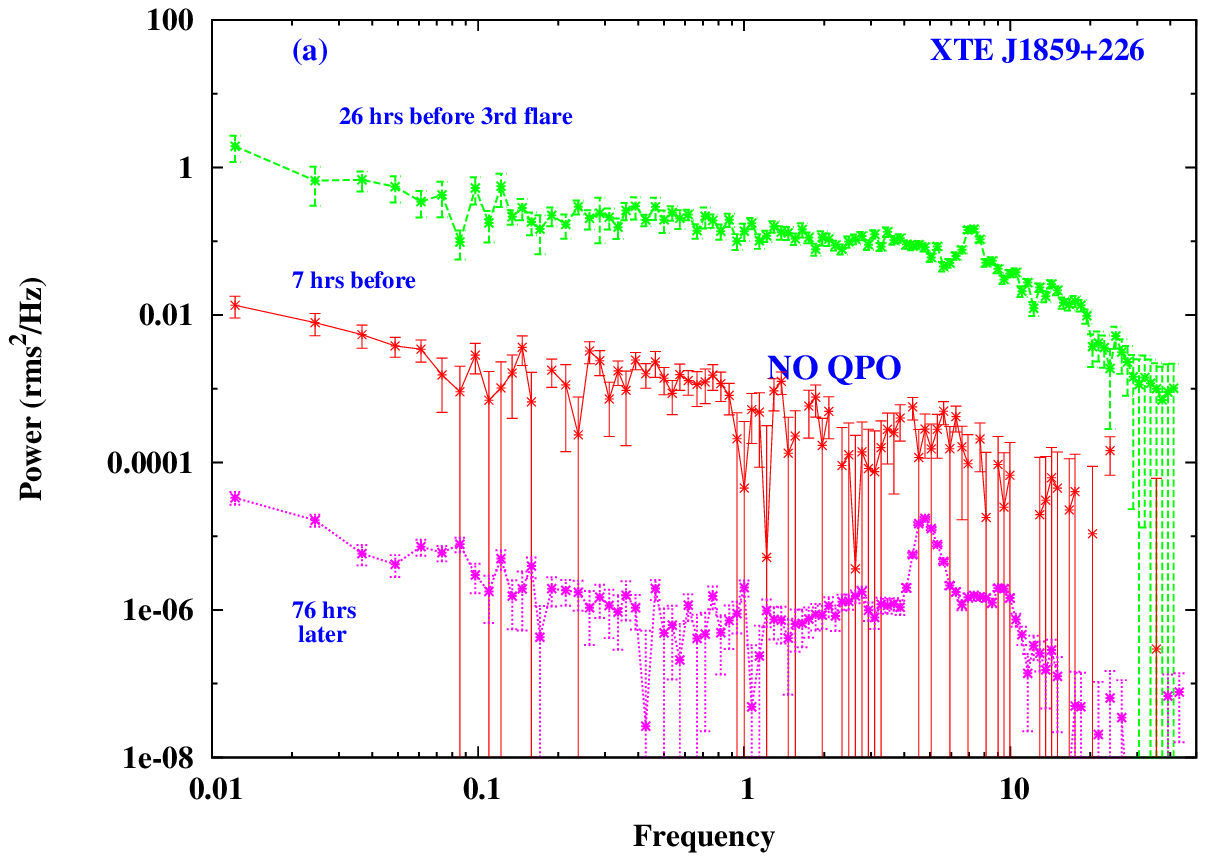}
\end{minipage}
\hfill
\begin{minipage}[t]{6cm}
\includegraphics[height=3cm,width=6cm]{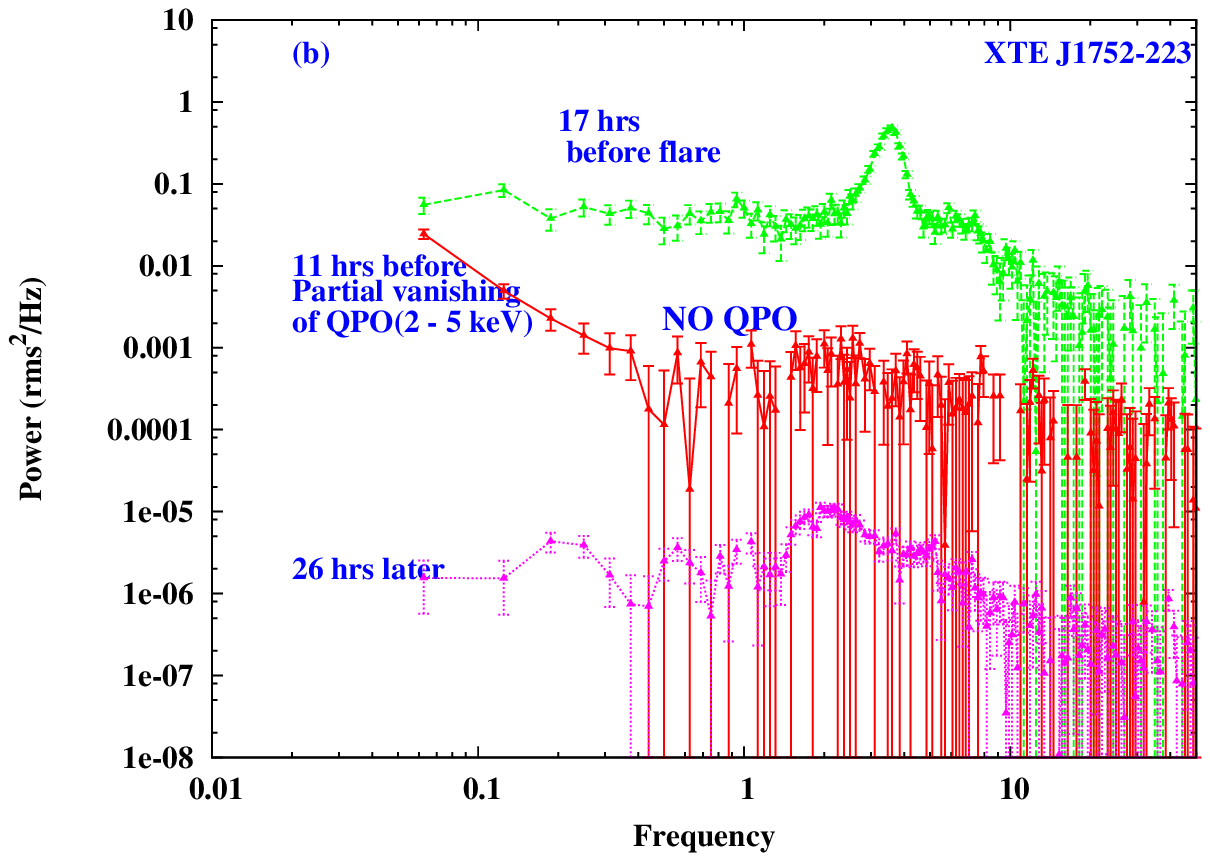}
\end{minipage}
\begin{minipage}[t]{6cm}
\includegraphics[height=3cm,width=6cm]{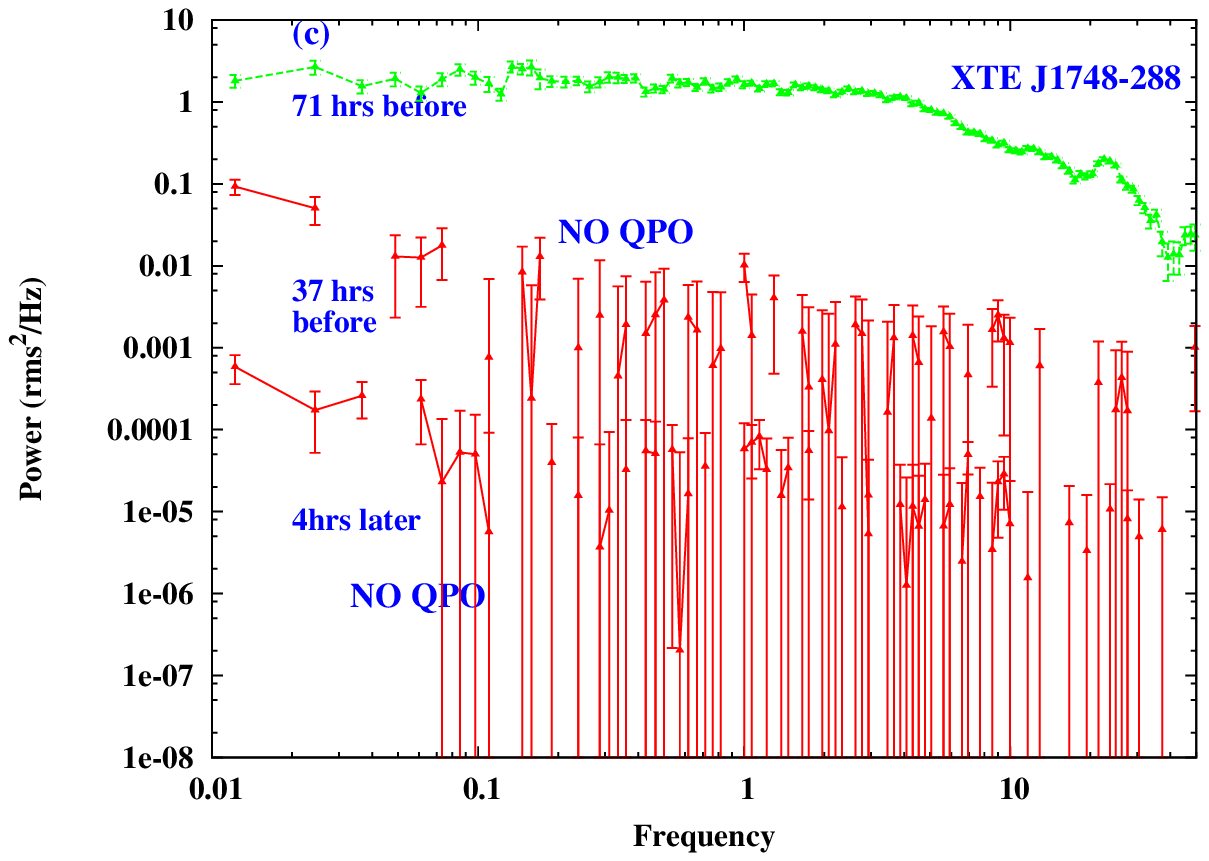}
\end{minipage}
\hfill
\begin{minipage}[t]{6cm}
\includegraphics[height=3cm,width=6cm]{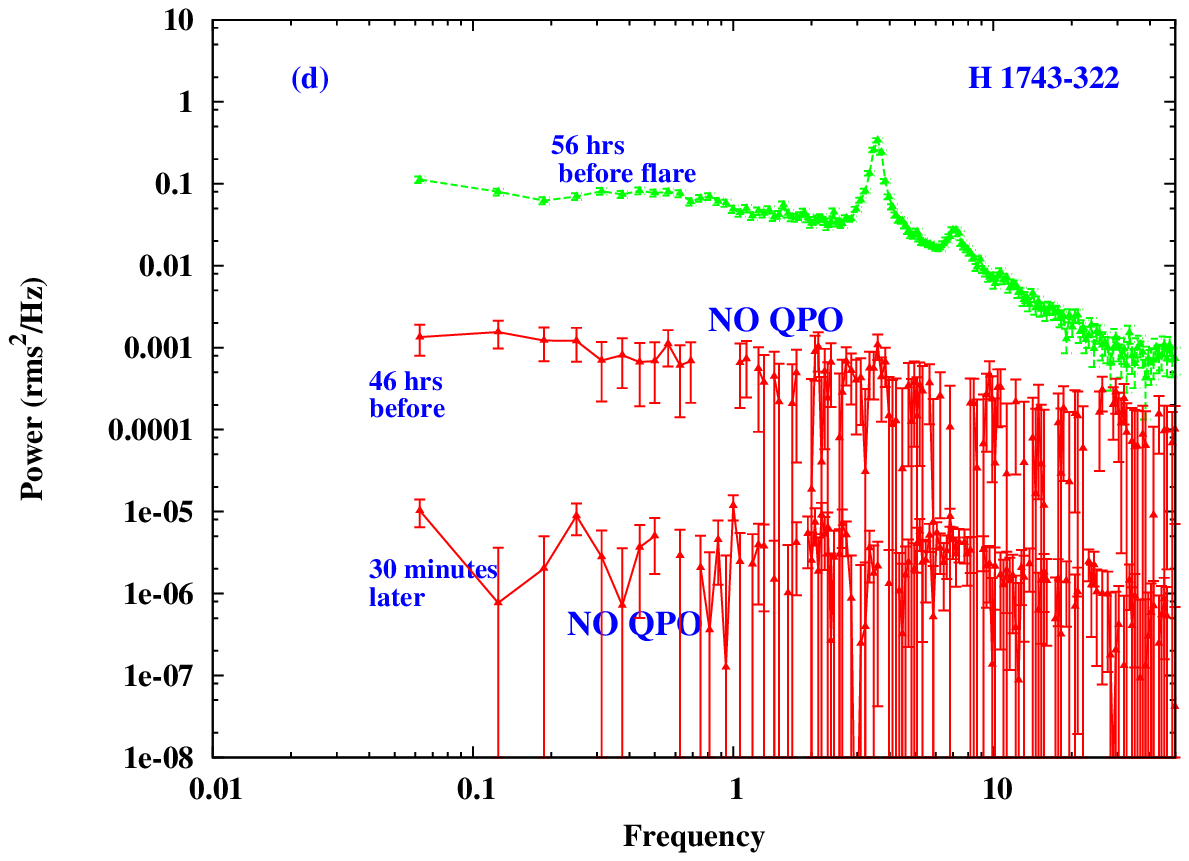}
\end{minipage}
\caption{PDS evolution observed during the flares indicating absence of QPOs (No signature of QPO in power spectra), for the sources 
XTE J1859$+$226, XTE J1752$-$223, XTE J1748$-$288 and H1743$-$322 (y-axis scaled in each plot 
for better representation).}
\label{figs}
\end{figure*}

\begin{itemize}
\item Partial disappearance of QPO (in 2 - 5 keV) during the first flare. Complete vanishing 
of QPOs during all the other flares (Panel a of Fig. \ref{figs}).
\item Possible detection of $6^{th}$ flare \citep{RA2012}.
\end{itemize}

\subsection{XTE J1752$-$223: 2009 outburst} 
\begin{itemize}
\item Partial vanishing of QPO in 2 - 5 keV 
\item Confirmation of the possible Jet ejection predicted in earlier observations 
(Panel b of Fig. \ref{figs}). 
\end{itemize}
\subsection{XTE J1748$-$288: 1998 outburst} 
\begin{itemize}
\item Disappearance of QPO, increase in spectral index, softening of spectra occurs 
(Panel c of Fig. \ref{figs}).
\item QPO disappearance \& spectral softening suggests presence of another flare, although no 
radio obs are available. 
\end{itemize}

\subsection{H 1743$-$322: 2009 outburst}
\begin{itemize}
\item Complete vanishing of QPO in 2 - 25 keV spectral band (Panel d of Fig. \ref{figs})
\item QPO reappears after 3 days of the prime ejection. 
\end{itemize}

\section{Conclusions}
The disk-jet connection observed in outbursting BH sources, can be understood based on the 
Magnetized-TCAF model \citep{AN2001}. Detailed results will be presented elsewhere 
(Radhika, Nandi \& Seetha, 2013, in prep.) and findings from the present work are 
summarized below:
\begin{itemize}
\item `Disappearance' of QPO occurs due to the `disruption' of inner part of the disk. 
\item Re-appearance of QPO after a few hours or days, suggests that the sub-Keplerian flow 
takes lesser time to refill the inner part of the disk.
\item No re-appearance of QPOs after a flare suggest that the inner part of the disk has been 
disrupted resulting into Jets, with a thermally dominated spectra.
\end{itemize}
 
\begin{acknowledgements}
We are thankful to Dr. P. Sreekumar of ISRO Satellite Centre for support related to 
participation in this conference and to Dr. T. Belloni for the usage of GHATS package.
\end{acknowledgements}

\bibliographystyle{aa}

\end{document}